\documentclass{iucrjournals}

\usepackage[utf8]{inputenc}
\usepackage{amsmath}
\usepackage{listings}
\usepackage{xcolor}
\usepackage{float}

\title{ASCRIBE-XR: Virtual Reality for Visualization of Scientific Imagery}

\author[a]{Ronald J. Pandolfi\IUCrCemaillink{ronpandolfi@lbl.gov}\IUCrOrcidlink{0000-0003-0824-8548}}%
\author[a]{Jeffrey J. Donatelli\IUCrCemaillink{jjdonatelli@lbl.gov}\IUCrOrcidlink{0009-0003-7173-0174
}}%
\author[d]{Julian Todd\IUCrEmaillink{julian@goatchurch.org.uk}}%
\author[a,b,c]{Daniela Ushizima\IUCrEmaillink{dushizima@lbl.gov}\IUCrOrcidlink{0000-0002-7363-9468}}%

\affil[a]{Applied Math and Computational Research Division, Lawrence Berkeley National Laboratory, Berkeley, CA, USA}
\affil[b]{Bakar Computational Health Sciences Institute, University of California San Francisco, San Francisco, CA, USA}
\affil[c]{Berkeley Institute for Data Science, University of California Berkeley, Berkeley, CA, USA}
\affil[d]{DoESLiverpool, Liverpool, England, UK}

\begin{document} 
\maketitle 

\begin{synopsis}
%One or two sentences suitable for the Journal contents listing and use in promoting your article via social media, highlighting the findings and significance of your work.
Discover ASCRIBE-XR, a novel PC-VR platform powered by Godot, available for Meta Quest headsets, enabling collaborative, real-time 3D visualization and exploration of science data to accelerate scientific discovery.
\end{synopsis}

\begin{abstract}
ASCRIBE-XR, a novel computational platform designed to facilitate the visualization and exploration of 3D volumetric data and mesh data in the context of synchrotron experiments, is described. Using Godot and PC-VR technologies, the platform enables users to dynamically load and manipulate 3D data sets to gain deeper insights into their research. The program's multi-user capabilities, enabled through WebRTC, and MQTT, allow multiple users to share data and visualize together in real-time, promoting a more interactive and engaging research experience. We describe the design and implementation of ASCRIBE-XR, highlighting its key features and capabilities. We will also discuss its utility in the context of synchrotron research, including examples of its application and potential benefits for the scientific community.
\end{abstract}


\section{Introduction}

The development of virtual environments for scientific purposes has become significantly more accessible, primarily due to three synergistic advancements. First, Extended Reality/Virtual Reality (XR/VR) software libraries have matured, offering more robust and abstract APIs that streamline complex interactions and rendering. Second, the seamless integration of these libraries into widely adopted game engines (like Unity, Unreal and Godot Engine) provides scientists with powerful, user-friendly development environments previously reserved for entertainment. Finally, the widespread availability of consumer VR hardware has driven down costs and increased the installed user base, making scientific VR applications more viable for broader dissemination and use. This confluence of factors enables scientists, for instance, to visualize diverse experimental data in an immersive VR environment, offering a deeper and more intuitive understanding of complex scientific contexts than traditional 2D visualizations. Recent efforts have shown progress and interest in applying consumer VR hardware to scientific data visualization \cite{Igarashi:vy5033,ZHANG2025106193,photonics12050436,Ushizima2025AscribeND}, which builds on substantial foundational research \cite{VANDAM2002535,Donalek,Marks,Korkut,Rubio}. 

\begin{figure}
    \centering
    \includegraphics[width=1\linewidth]{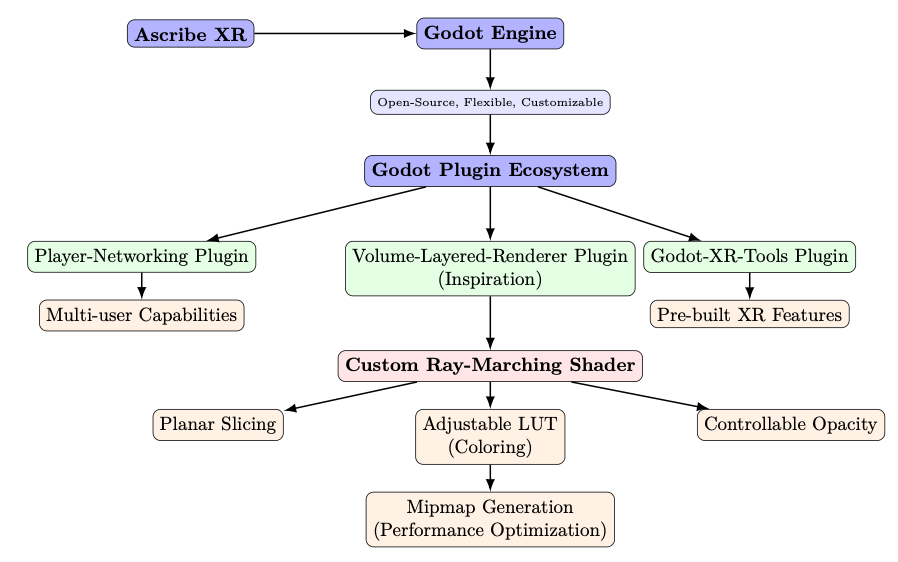}
        \caption{Architectural Overview of Ascribe XR's Godot Integration and Features}
    \label{fig:ascribe_xr_architecture}
\end{figure}

In this context, we introduce ASCRIBE-XR (Figure~\ref{fig:ascribe_xr_architecture}), a novel computational platform designed to facilitate the visualization and exploration of 3D volumetric and mesh data from scientific experiments, such as synchrotron data, electron microscopy and more. ASCRIBE-XR, written in Godot, provides an immersive and interactive platform for scientists and engineers to analyze and interpret their data in a virtual reality (VR) environment. By leveraging the capabilities of Godot and PC-VR technology, ASCRIBE-XR enables users to dynamically load and manipulate 3D data sets to gain deeper insight into their research. The program's ability to handle diverse data makes it an ideal tool for synchrotron imaging techniques such as tomography.

\begin{figure}[h]
    \centering
    \includegraphics[width=0.75\linewidth]{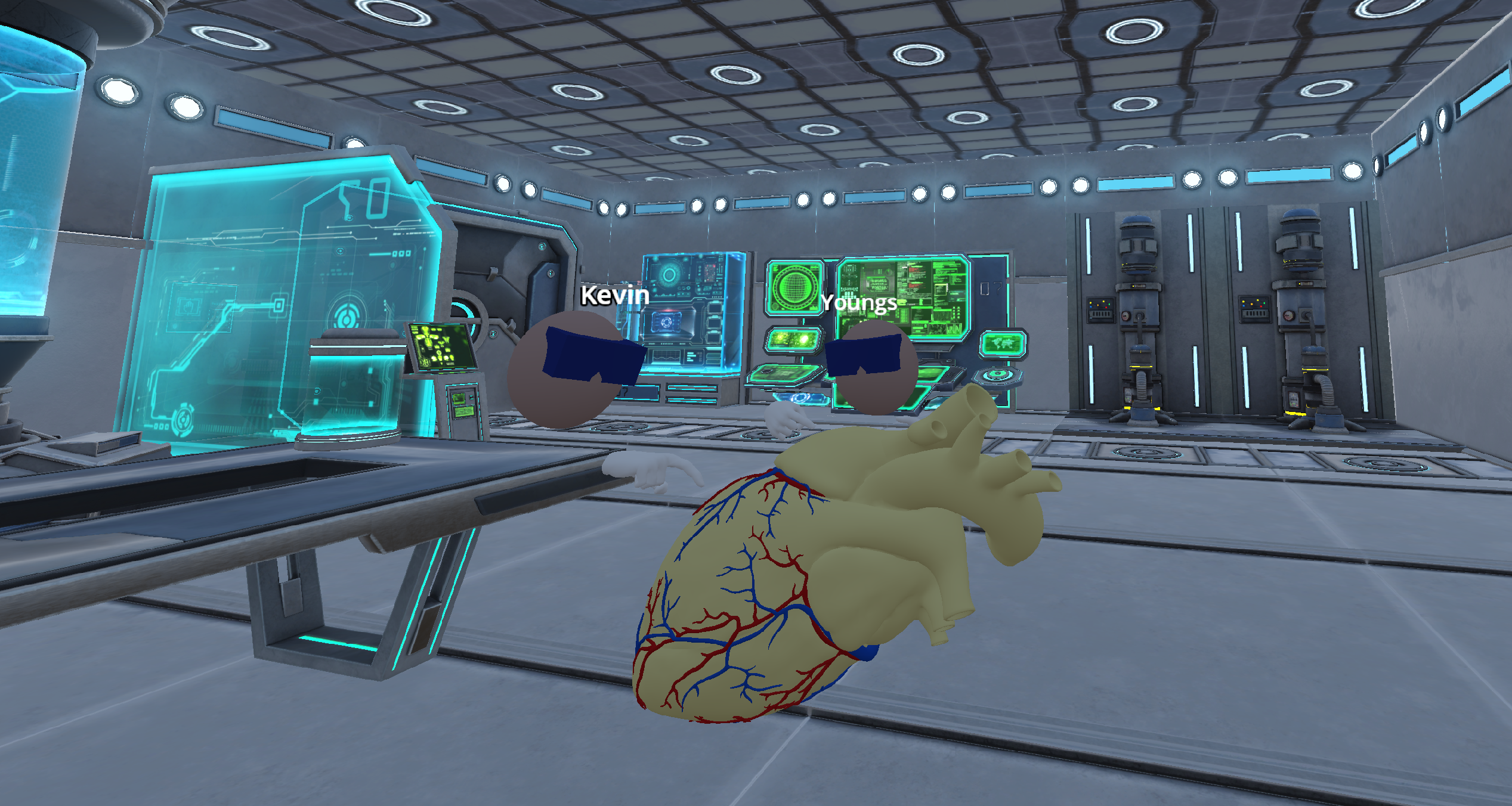}
    \caption{Collaborative research in action: snapshot illustrates ASCRIBE-XR's multi-user environment, where researchers share and visualize data together in real-time. This interactive experience is facilitated by WebRTC and MQTT for data synchronization, complemented by VOIP with Opus compression for clear voice communication within the VR laboratory.}
    \label{fig:multiuser}
\end{figure}

The use of VR technology in ASCRIBE-XR offers a unique advantage in the analysis and interpretation of complex 3D data sets. By immersing users in a virtual environment, ASCRIBE-XR enables researchers to explore and interact with their data in a more intuitive and engaging way, potentially leading to new insights and discoveries. One of the key features of ASCRIBE-XR is its ability to facilitate collaboration and data sharing among researchers. Figure ~\ref{fig:multiuser} show ASCRIBE-XR's multi-user capabilities, enabled through WebRTC and MQTT, allow multiple users to share data and visualize together in real-time, promoting a more interactive and engaging research experience. Additionally, the program's support for voice communication through VOIP using Opus compression enables seamless communication among collaborators.

In this article, we describe the design and implementation of ASCRIBE-XR, highlighting its key features and capabilities. We will also discuss its utility in the context of synchrotron research, including examples of its application and potential benefits for the scientific community.

\section{Design and Implementation}

ASCRIBE-XR is built using the Godot game engine, a popular open-source engine known for its flexibility, customizability, and large community of developers. Godot's architecture and extensive plugin ecosystem make it an ideal choice for building complex, interactive applications like ASCRIBE-XR. Specifically, Godot's support for XR and its large collection of community-developed plugins enabled rapid development and prototyping of ASCRIBE-XR's core features. Godot Engine utilizes OpenXR as its core standard for XR (VR/AR/MR) development, offering a unified API for interacting with various XR platforms.

In the case of ASCRIBE-XR, we made extensive use of community plugins with compartmentalized functionality to implement key features. For example, the player-networking plugin enabled the development of ASCRIBE-XR's multi-user capabilities, allowing multiple users to share data and visualize together in real-time. The \texttt{openxrvendors} plugin provided a standardized interface for interacting with XR devices, while the \texttt{godot-xr-tools} plugin offered a set of pre-built XR-related features and tools. To handle 3D volumetric data, we utilized the \texttt{volume-layered-renderer} plugin as a starting point and adapted it with our own graphics shader since the original plugin uses a layer-based rendering approach. We additionally implemented our own ray-marching-based shader to provide better quality rendering of volumetric data.

Our custom shader supports exclusion planes for planar slicing, allowing users to inspect the interior of a volume more clearly. Additionally, the shader uses an adjustable Look-Up Table (LUT) for coloring, enabling users to customize the color mapping of the volume data. The shader also provides controllable opacity, allowing users to adjust the transparency of the volume. The use of a ray-marching-based shader generally provides several benefits, including reduced memory usage and upgraded quality. The adjustable LUT and controllable opacity also provide users with a high degree of control over the visualization of their data, allowing them to customize the appearance of the volume to suit their needs.

To optimize rendering performance, we generate mipmaps for the 3D texture data when it is loaded. This technique improves rendering at a distance, reducing the amount of data that need to be rendered, and enhancing overall performance.

By leveraging Godot's plugin ecosystem and building a modular, extensible architecture, we were able to rapidly develop and prototype ASCRIBE-XR's core features, and create a powerful and flexible tool for scientists and engineers working with 3D data sets.

%Commented only for arxiv 
%Appendix table \ref{tab:godot_components} provides a categorized overview of these dependencies, detailing their function and source. This table serves as a high-level architectural map, illustrating the functional layers of the technology stack, from core immersive functionalities and networking to world generation and developer tooling.

\section{Using ASCRIBE-XR}

ASCRIBE-XR is designed to be a user-friendly and accessible tool for scientists and engineers working with 3D data sets. The application uses OpenXR to abstract different VR headsets, allowing users to run ASCRIBE-XR on a variety of devices. However, pass-through capabilities are currently only supported on Meta Quest devices through a vendor-specific plugin.

\begin{figure}[t] %
\label{fig:cmc}
\begin{center}
\includegraphics[width=0.5\textwidth]{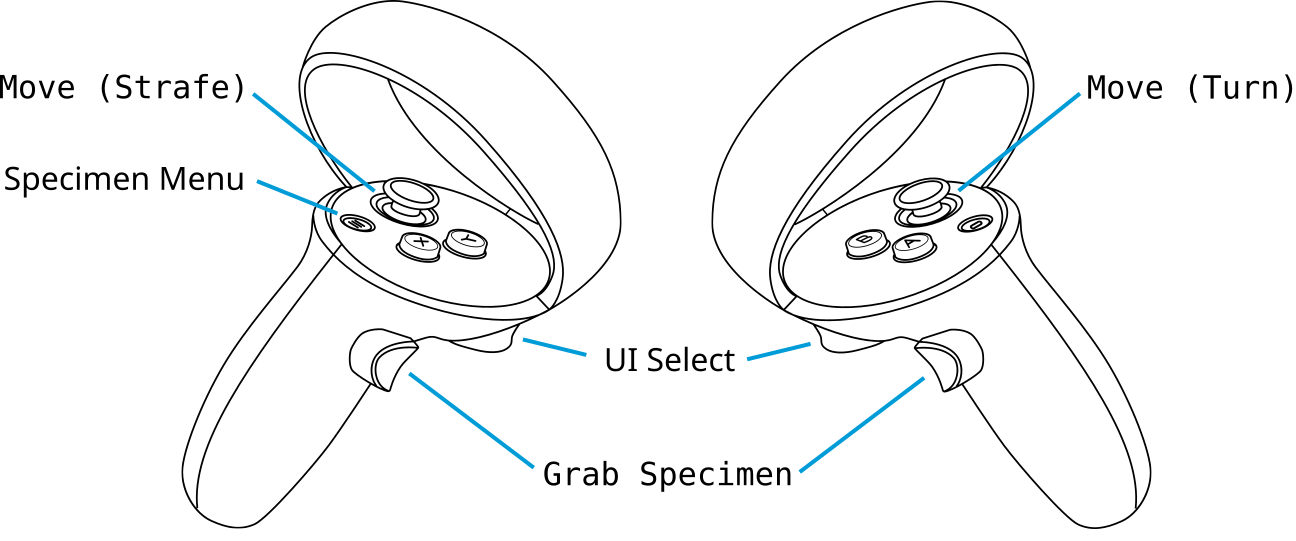} % NB use pdflatex for non-postscript
\end{center}
\caption{Ascribe-XR controller button scheme, shown for Quest Touch controllers.} % NB \protect\cite{...} is required in floating figures
\end{figure}

ASCRIBE-XR can be run standalone or with PC-VR, but for visualizing large volumetric data, running with PC-VR provides better performance and quality. To use ASCRIBE-XR with PC-VR, users connect their headset to their PC and activate a link application such as Meta Link, SteamVR/Steam Link, or Virtual Desktop. Then, they can run the ASCRIBE-XR executable and users are loaded into a virtual scene with a menu of ``Specimens'' to select from.

To interact with the application, users can use the following controls:

\begin{itemize}
    \item Trigger buttons: used to ``click'' on specimen buttons to load a specimen;
    \item Grip buttons: grab and manipulate specimens;
    \item Left joystick: moves the user relative to their facing;
    \item Right joystick: turns the user's facing left/right.
\end{itemize}

From the main menu, users can select a specimen to load by ``clicking' on a button. Once a specimen is loaded, users can manipulate it by grabbing it with the grip buttons. The user can also load their own 3D data sets using the ``Load Mesh'' and ``Load Volume'' options.

The ``Load Mesh'' option allows users to select a file in one of the following formats: glTF 2.0, Blender, DAE, OBJ, FBX, or STL. The selected mesh is then dynamically imported as a specimen (geometry) mesh. Similarly, the ``Load Volume'' option allows users to select a file in one of the following formats: zipped image stacks, .npy, or .bin. The selected volume is then dynamically imported as a specimen (dense matrix) volume.

In both cases, a user interface is shown for adjusting the visualization, including controls for:

\begin{itemize}
    \item Scale: adjust the size of the specimen;
    \item Material shader: select a material shader to apply to the specimen;
    \item LUT: adjust the color mapping of the specimen;
    \item Opacity: adjust the transparency of the specimen;
    \item Render quality: adjust rendering quality (options vary by shader).
\end{itemize}

Pressing the ``Menu'' button brings the main menu back, allowing users to load a different specimen or exit the application.

By providing an intuitive and user-friendly interface, ASCRIBE-XR enables scientists and engineers to easily explore and analyze their 3D data sets in an immersive and interactive environment as illustrate in Figures~\ref{fig:concrete}-~\ref{fig:virus}.

\section{Networking Architecture}

For a scientific VR application to maximize its collaborative reach and facilitate effective expert-to-novice knowledge transfer, a robust multiplayer feature is invaluable. This enables simultaneous ``in-world'' interaction, allowing experts to directly demonstrate complex functionalities and data interpretations to new users. While the foundational networking for such a feature has conventionally relied on client applications connecting to a static domain-provisioned server, modern VR multiplayer solutions also critically address challenges like maintaining ultra-low latency for immersive presence, synchronizing intricate physical interactions, and optimizing data replication across distributed clients. Unfortunately, the maintenance of this architecture (setting up accounts, deploying server software, distributing login credentials, etc.) is normally prohibitive for small rough-and-ready projects -- exactly the type that most need to be expertly demonstrated.  Even with special online services (such as https://www.gd-sync.com/ ) to streamline the process, it can still be a hassle.

WebRTC is a protocol that establishes a direct network link (peer to peer) between two computers connected anywhere on the internet without the need of intervening server hosted in a fixed domain. As such it is ideal for linking two computers hosting VR instances without the need of any additional systems. The only requirement is that each computer must exchange its current location on the internet (often behind various routers) with its peers to connect to each another. For this purpose we found that the MQTT protocol (which was designed for transmitting IoT sensor information) provides a simple and transparent system for sharing state and small packets of information between computers that are unknown to one another.  Using this method, it is possible to make the multiplayer feature sufficiently low-maintenance that it can be used to gain feedback from non-experts.
 
\section{Example Applications using Diverse Image Modalities}
\begin{figure}[t]
    \centering
    \includegraphics[width=0.75\linewidth]{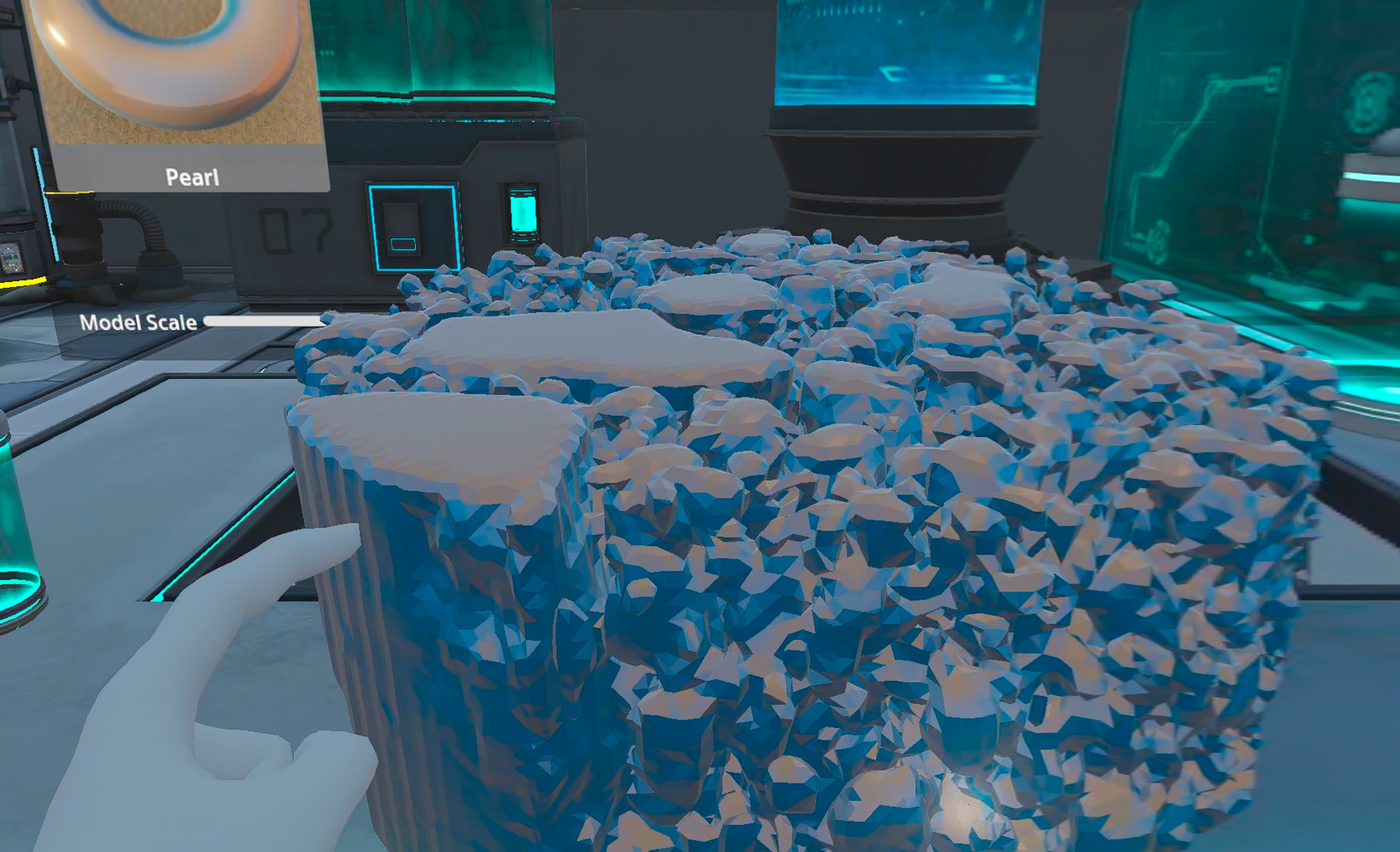}
    \caption{Sample of archaeological concrete from Pompeii, Italy shown (left) in a sequence of image slices and (right) rendered in ASCRIBE-XR with with ``Pearl'' material.}
    \label{fig:concrete}
\end{figure}

Our prior work in semantic segmentation has provided a strong basis, encompassing a range of techniques applied across diverse image modalities. This includes contributions to unsupervised segmentation, such as utilizing superpixel representations with Voronoi diagrams~\cite{ISBI:2015}. We have also advanced supervised methods, employing deep learning approaches with U-Net, Residual U-Net, and transformer architectures \cite{Brain:2017,desiqueiraUshizima2022,Battery:NPJ:2023,Quenum:2023} for robust and accurate segmentation. Following one or more segmentation schemes to create binary partitions of 3D images in Figures~\ref{fig:concrete}-~\ref{fig:cmc}, we run marching cubes to extract the mesh of the external layer of solids, enabling further analysis and visualization, as discussed in this section.

%X
%microCT
\subsection{Archaeological concrete}
The exceptional durability of ancient Roman concrete, its low carbon footprint and unique interlocking mineral structure offer valuable insight for designing advanced materials. Lawrence Berkeley National Laboratory (LBNL) hosts a variety of instruments exploiting different light sources, including X-ray tomography (XRT) at the Advanced Light Source (ALS), often used to investigate the structural characteristics of materials, such as archaeological samples ~\cite{Xu:2023,MRS:2020:concrete}.

As discussed in ~\cite{MRS:2020:concrete,BadranUshizima2022}, our materials data science algorithms enable analysis of XRT images to inspect and quantify the microstructure of ancient Roman concrete samples. We use XRT data to define material phases, estimate fractions, and visualize porous networks and density gradients within a remarkable sample, ocean-submersed for two millennia, which far outlast modern concrete. Their internal organization appears to be key to their incredible longevity.

Previous results highlight the advantage of combining non-destructive 3D XRT with computer vision and machine learning for characterizing complex archaeological materials. A major impact of this work was its ability to significantly reduce computational data, allowing for near real-time analysis on minimal infrastructure, potentially at the imaging facility. This refined data then serves as the input for ASCRIBE-XR, our visualization tool, which brings these intricate 3D microstructures to life.

%fibers = microCT
\subsection{Ceramic matrix composites}
Ceramic Matrix Composites (CMCs) are next-generation materials designed for extreme temperatures and deformations in aerospace and industrial turbine applications. CMCs outperform nickel super-alloys by operating up to $500^\circ\text{F}$ hotter and weighing one-third less, leading to lighter, more efficient engines and reduced emissions. MicroCT volumes are key for non-destructively characterizing CMC 3D microstructure.

Previous work~\cite{parkinson2017machine,SC:2020:Dani} used microCT to investigate CMC microstructure evolution during manufacturing processes like polymer-to-ceramic conversion and fiber impregnation, vital for reinforcement. Those samples typically comprised over a thousand tows of Hi-Nicalon silicon carbide (SiC) fibers, each with about 500 fibers averaging $6.4 \pm 0.9 \, \mu\text{m}$ in radius. These fiber beds were approximately 1.5mm wide, containing 5,000-6,200 fibers.

\begin{figure}[th] %
\label{fig:cmc}
\begin{center}
\includegraphics[width=0.75\textwidth]{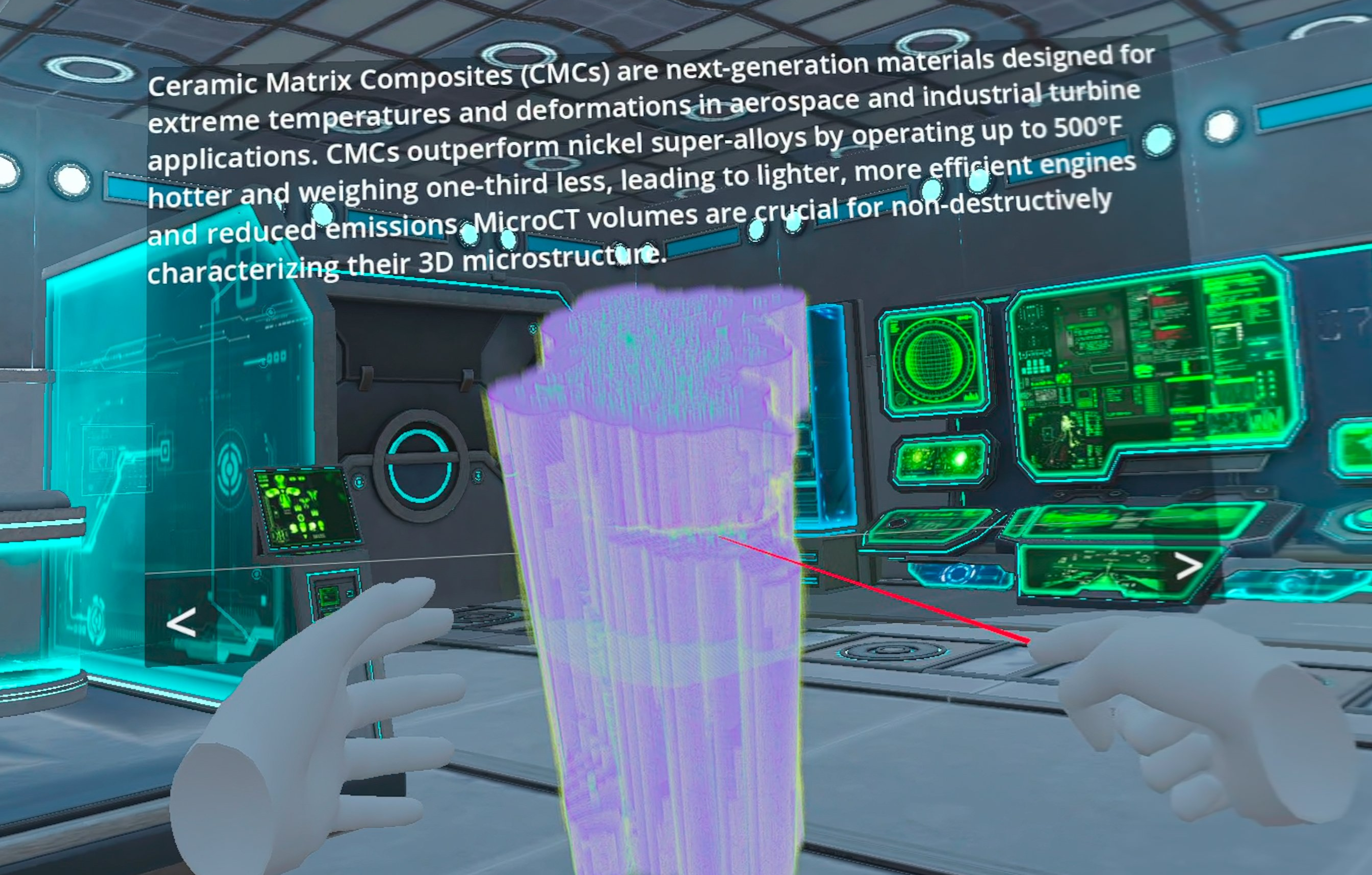} % NB use pdflatex for non-postscript
\end{center}
\caption{A ceramic matrix composite specimen rendered with the ray-marching shader. A browsable description of the context of the specimen is shown in the background. This shader has many tunable properties allowing the visualization to be adjusted.} % NB \protect\cite{...} is required in floating figures
\end{figure}

Public data repositories have been essential for algorithm development and reproducibility~\cite{Siqueira_van_der_Walt_Ushizima_2021}. Those microCT images, each about 60 GB and over 14 billion voxels, were acquired at the LBNL ALS and are available via the Materials Data Facility~\cite{Blaiszik_2016}. This paper utilizes one stack of these public volumes,.

\subsection{Virus}

\begin{figure}[ht]
    \centering
    \includegraphics[width=0.75\linewidth]{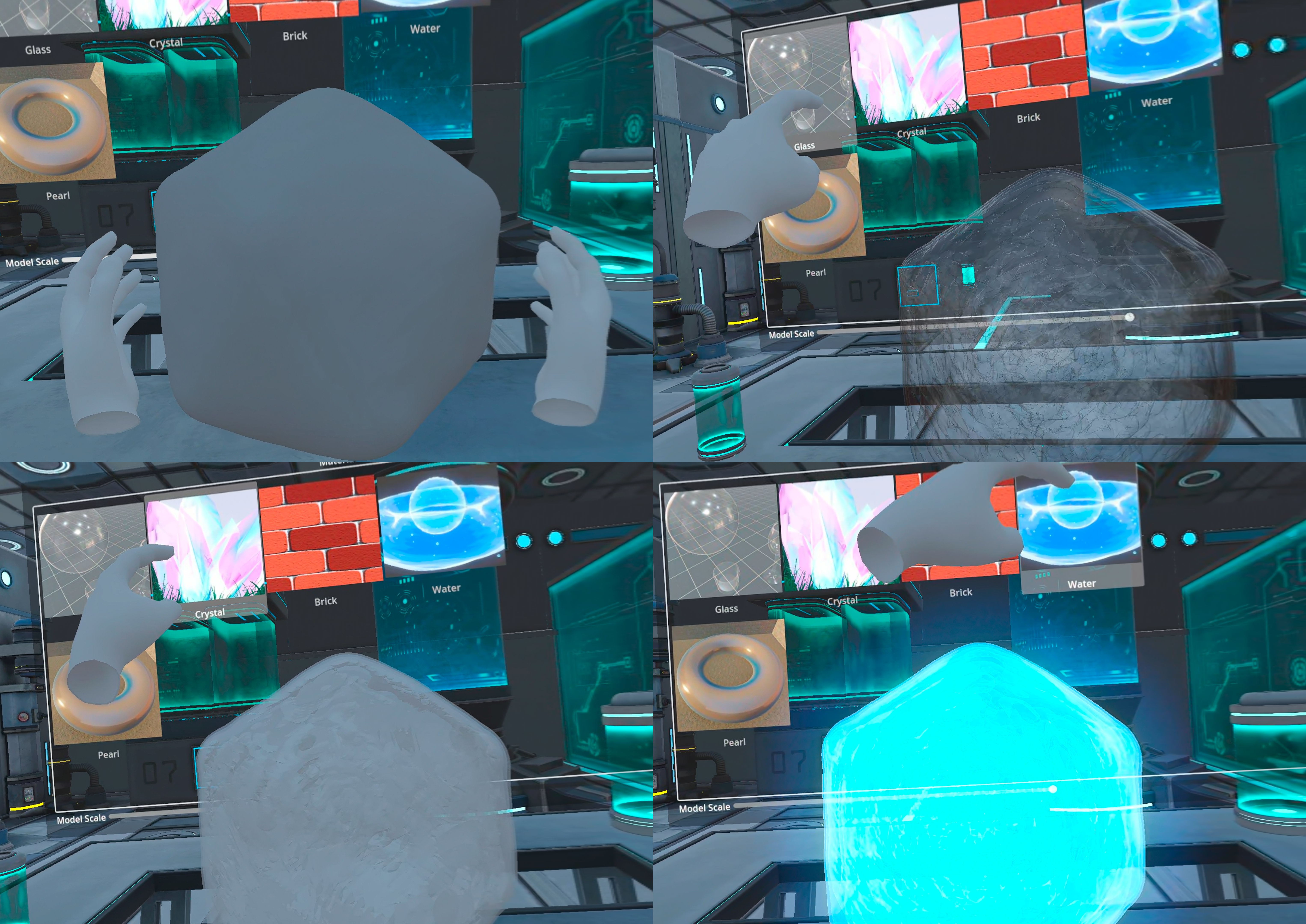}
    \caption{Visualizations of the PBCV-1 virus (clockwise from top left) demonstrate the impact of different shader materials: ``default gray'', ``glass'', ``water'', and ``crystal''. A library of these materials allows users to highlight various structural features. Additionally, some shaders offer tunable properties for fine-grained control.}
    \label{fig:virus}
\end{figure}

Figure~\ref{fig:virus} presents a compelling 3D reconstruction of the Paramecium bursaria chlorella virus 1 (PBCV-1) \cite{doi:10.1073/pnas.1812064115,Donatelli2015Iterative}, highlighting a significant advance in structural biology and imaging. This large, double-stranded DNA virus, known for infecting Chlorella-like algae, is a prime candidate for high-resolution studies due to its sheer size (175-190 nm) and intricate nature. PBCV-1 is defined by its icosahedral shape and multi-layered composition. Its outer glycoprotein capsid shell is precisely built from capsomers in a T=169d triangulation, including a distinct spike structure at one vertex that likely facilitates host cell binding. Encased within an internal lipid bilayer membrane, its large, double-stranded DNA genome resides. 

The detail presented in this figure, allowing us to visualize the virus's external structure, its intricate capsid protein, and even its internal viral proteins, is made possible through cutting-edge technology. Specifically, these reconstructions are meticulously derived from fluctuation X-ray scattering data, only possible through the CAMERA M-TIP framework~\cite{Donatelli2015Iterative}. This specialized technique allows researchers to capture subtle variations in how X-rays scatter from individual virus particles, providing the raw information needed to build a comprehensive 3D model.

This high-quality data was collected at one of the world's premier scientific facilities: the Linac Coherent Light Source (LCLS) at SLAC National Accelerator Laboratory. The LCLS is a groundbreaking hard X-ray free-electron laser, capable of producing X-ray pulses that are not only incredibly bright but also ultrashort, on the femtosecond timescale. This unique capability allows scientists to essentially take "snapshots" of atoms and molecules in action, capturing dynamic processes that were previously unobservable. For PBCV-1, this means probing its structure with unprecedented clarity, revealing the fundamental building blocks and arrangements that dictate its biological function. The synergy between advanced experimental techniques at facilities like LCLS, and sophisticated reconstruction algorithms, such as those employed by ASCRIBE-XR, truly unlocks new frontiers in our understanding of complex biological systems.

\section{Conclusion}

In conclusion, ASCRIBE-XR represents a significant advancement in the field of scientific data visualization, offering an immersive and interactive platform for scientists and engineers to analyze and interpret complex 3D data sets. By leveraging the capabilities of Godot and PC-VR technology, ASCRIBE-XR enables users to dynamically load and manipulate 3D data sets, facilitating a deeper understanding of their research. The program's ability to handle diverse data, multi-user capabilities, and support for voice communication make it an ideal tool for collaborative research.

The applications of ASCRIBE-XR are diverse. The program's utility in shown examples highlights its potential to facilitate new insights and discoveries across various scientific disciplines. Furthermore, ASCRIBE-XR's user-friendly interface and accessibility make it an attractive tool for researchers, educators, and students.

ASCRIBE-XR has the potential to transform the way scientists and engineers interact with their data, enabling a more intuitive and engaging research experience. Its development and application demonstrate the power of interdisciplinary collaboration and the importance of innovative technologies in advancing scientific understanding. Future work will focus on continued development and refinement of ASCRIBE-XR to integrate dynamic content with live data processing.

%Commented only for arxiv  
%\input{appendix}     

\begin{acknowledgements}
This work was supported by the US Department of Energy (DOE) Office of Science Advanced Scientific Computing Research (ASCR) and Basic Energy Sciences (BES) under Contract No. DE-AC02-05CH11231 to the Center for Advanced Mathematics for Energy Research Applications (CAMERA) program. It also included support from the DOE ASCR-funded project Autonomous Solutions for Computational Research with Immersive Browsing \& Exploration (ASCRIBE), which is supported by the Office of Science of the U.S. Department of Energy under Contract No. DE-FOA-0003545.
\end{acknowledgements}

% \begin{funding}
% List funding organizations, recipients, grant numbers, etc.
% \end{funding}

% \ConflictsOfInterest{Please declare any conflicts of interest, or declare  that there are no conflicts of interest.
% }

% \DataAvailability{Please state how the data supporting the results reported in your article can be accessed, e.g. within the article, as published supporting material, in repositories, upon request...
% }

\bibliography{library} % basename of .bib file

@article{Donatelli2015Iterative,
  title     = {Iterative phasing for fluctuation X-ray scattering},
  author    = {Donatelli, Jeffrey J. and Zwart, Peter H. and Sethian, James A.},
  journal   = {Proceedings of the National Academy of Sciences},
  volume    = {112},
  number    = {33},
  pages     = {10286--10291},
  year      = {2015},
  doi       = {10.1073/pnas.1513738112},
  url       = {https://doi.org/10.1073/pnas.1513738112},
}

@article{Igarashi:vy5033,
author = "Igarashi, Haruo and Kido, Daiki and Ishii, Yutaka and Niwa, Yasuhiro and Okamoto, Atsushi and Kimura, Masao",
title = "{Visualization of four-dimensional X-ray absorption fine structure data using a virtual reality system}",
journal = "Journal of Synchrotron Radiation",
year = "2025",
volume = "32",
number = "1",
pages = "162--170",
month = "Jan",
doi = {10.1107/S1600577524011226},
url = {https://doi.org/10.1107/S1600577524011226},
}

@article{VANDAM2002535,
title = {Experiments in Immersive Virtual Reality for Scientific Visualization},
journal = {Computers \& Graphics},
volume = {26},
number = {4},
pages = {535-555},
year = {2002},
issn = {0097-8493},
doi = {https://doi.org/10.1016/S0097-8493(02)00113-9},
url = {https://www.sciencedirect.com/science/article/pii/S0097849302001139},
author = {Andries {van Dam} and David H Laidlaw and Rosemary Michelle Simpson}
}

@INPROCEEDINGS{Donalek,
  author={Donalek, Ciro and Djorgovski, S. G. and Cioc, Alex and Wang, Anwell and Zhang, Jerry and Lawler, Elizabeth and Yeh, Stacy and Mahabal, Ashish and Graham, Matthew and Drake, Andrew and Davidoff, Scott and Norris, Jeffrey S. and Longo, Giuseppe},
  booktitle={2014 IEEE International Conference on Big Data (Big Data)}, 
  title={Immersive and collaborative data visualization using virtual reality platforms}, 
  year={2014},
  volume={},
  number={},
  pages={609-614},
  doi={10.1109/BigData.2014.7004282}}

@inproceedings{Marks,
author = {Marks, Stefan and Estevez, Javier E. and Connor, Andy M.},
title = {Towards the Holodeck: Fully Immersive Virtual Reality Visualisation of Scientific and Engineering Data},
year = {2014},
isbn = {9781450331845},
publisher = {Association for Computing Machinery},
address = {New York, NY, USA},
url = {https://doi.org/10.1145/2683405.2683424},
doi = {10.1145/2683405.2683424},
booktitle = {Proceedings of the 29th International Conference on Image and Vision Computing New Zealand},
pages = {42–47},
numpages = {6},
location = {Hamilton, New Zealand},
series = {IVCNZ '14}
}

@article{Korkut,
author = {Korkut, Elif Hilal and Surer, Elif},
title = {Visualization in virtual reality: a systematic review},
year = {2023},
issue_date = {Jun 2023},
publisher = {Springer-Verlag},
address = {Berlin, Heidelberg},
volume = {27},
number = {2},
issn = {1359-4338},
url = {https://doi.org/10.1007/s10055-023-00753-8},
doi = {10.1007/s10055-023-00753-8},
journal = {Virtual Real.},
month = jan,
pages = {1447–1480},
numpages = {34}
}

@article{ZHANG2025106193,
title = {XR-based interactive visualization platform for real-time exploring dynamic earth science data},
journal = {Environmental Modelling \& Software},
volume = {183},
pages = {106193},
year = {2025},
issn = {1364-8152},
doi = {https://doi.org/10.1016/j.envsoft.2024.106193},
url = {https://www.sciencedirect.com/science/article/pii/S1364815224002548},
author = {Xuelei Zhang and Hu Yang and Chunhua Liu and Qingqing Tong and Aijun Xiu and Lingsheng Kong and Mo Dan and Chao Gao and Meng Gao and Huizheng Che and Xin Wang and Guangjian Wu}
}

@Article{photonics12050436,
AUTHOR = {Matys, Martin and Thistlewood, James P. and Kecová, Mariana and Valenta, Petr and Greplová Žáková, Martina and Jirka, Martin and Hadjisolomou, Prokopis and Špádová, Alžběta and Lamač, Marcel and Bulanov, Sergei V.},
TITLE = {Visualization of High-Intensity Laser–Matter Interactions in Virtual Reality and Web Browser},
JOURNAL = {Photonics},
VOLUME = {12},
YEAR = {2025},
NUMBER = {5},
ARTICLE-NUMBER = {436},
URL = {https://www.mdpi.com/2304-6732/12/5/436},
ISSN = {2304-6732},
DOI = {10.3390/photonics12050436}
}

@article{Rubio,
author = {Rubio-Tamayo, Jose Luis and Wuebben, Daniel and Gértrudix, Manuel},
year = {2024},
month = {04},
pages = {},
title = {Standards for science communication in extended and virtual reality: a model for XR/VR based on London Charter and Seville Principles},
volume = {23},
journal = {Journal of Science Communication},
doi = {10.22323/2.23030203}
}

@article{ISBI:2015,
author = {G. L. B. Ramalho and D. S. Ferreira and A. G. C. Bianchi and C. M. Carneiro and F. N. S. Medeiros and D. M. Ushizima},
title ={Cell reconstruction under {Voronoi} and enclosing ellipses from 3D microscopy},
journal = {{IEEE International Symposium on Biomedical Imaging (ISBI)}},
month = apr,
address = "New York, NY",
year= {2015},
}

@article{Xu:2023,
title = {In-situ microtomography image segmentation for characterizing strain-hardening cementitious composites under tension using machine learning},
journal = {Cement and Concrete Research},
volume = {169},
pages = {107164},
year = {2023},
issn = {0008-8846},
doi = {https://doi.org/10.1016/j.cemconres.2023.107164},
url = {https://www.sciencedirect.com/science/article/pii/S0008884623000789},
author = {Ke Xu and Qingxu Jin and Jiaqi Li and Daniela M. Ushizima and Victor C. Li and Kimberly E. Kurtis and Paulo J.M. Monteiro}}

@inproceedings{parkinson2017machine,
  title={Machine learning for micro-tomography},
  author={Parkinson, Dilworth Y and Pelt, Dani{\"e}l M and Perciano, Talita and Ushizima, Daniela and Krishnan, Harinarayan and Barnard, Harold S and MacDowell, Alastair A and Sethian, James},
  booktitle={Developments in X-Ray Tomography XI},
  volume={10391},
  pages={103910J},
  year={2017},
  organization={International Society for Optics and Photonics}
}

@article{Brain:2017,
  author    = {Maryana Alegro and
               Panagiotis Theofilas and Austin Nguy and
               Patricia A. Castruita and William Seeley
               and Helmut Heinsen and Daniela Ushizima and Lea T. Grinberg},
  title     = {Automating Cell Detection and Classification in Human
        Brain Fluorescent Microscopy Images Using Dictionary
        Learning and Sparse Coding},
  journal = {Journal of Neuroscience Methods},
  pages     = {20--33},
  volume  = {282},
  year      = {2017},

}

@article{Battery:NPJ:2023,
	author = {Huang, Ying and Perlmutter, David and Su, Andrea and Quenum, Jerome and Zenyuk, Iryna and Ushizima, Daniela},
	isbn = {2057-3960},
	journal = {Nature Partner Journal {(NPJ)} Computational Materials},
	title = {Detecting Lithium Plating Dynamics in a Solid-State Battery with Operando X-ray Computed Tomography using Machine Learning (accepted)},
	year = {2023}
}

@article{MRS:2020:concrete,
title =  "Materials Data Science for Microstructural Characterization of Archaeological Concrete",
author = "Daniela Ushizima and Ke Xu and Paulo Monteiro",
journal = "MRS Advancements - special issue: Materials Data Science",
pages = "1-14",
year = "2020",
publisher = "Cambridge Press",
}

@Article{BadranUshizima2022,
AUTHOR = {Badran, Aly and Parkinson, Dula and Ushizima, Daniela and Marshall, David and Maillet, Emmanuel},
TITLE = {Validation of Deep Learning Segmentation of CT Images of Fiber-Reinforced Composites},
JOURNAL = {Journal of Composites Science},
VOLUME = {6},
YEAR = {2022},
NUMBER = {2},
ARTICLE-NUMBER = {60},
URL = {https://www.mdpi.com/2504-477X/6/2/60},
ISSN = {2504-477X},
DOI = {10.3390/jcs6020060}
}

@article{desiqueiraUshizima2022,
	author = {Fioravante de Siqueira, Alexandre and Ushizima, Daniela M. and van der Walt, St{\'e}fan J.},
	da = {2022/02/02},
	doi = {10.1038/s41597-022-01119-6},
	id = {Fioravante de Siqueira2022},
	isbn = {2052-4463},
	journal = {Scientific Data},
	number = {1},
	pages = {32},
	title = {A reusable neural network pipeline for unidirectional fiber segmentation},
	ty = {JOUR},
	url = {https://doi.org/10.1038/s41597-022-01119-6},
	volume = {9},
	year = {2022}}

@InProceedings{SC:2020:Dani,
author = {Daniela Ushizima and Matthew McCormick and Dilworth Parkinson},
title = {Accelerating Microstructural Analytics with Dask for Volumetric X-ray Images},
booktitle = {2020 IEEE/ACM 9th Workshop on Python for High-Performance and Scientific Computing (PyHPC) at Super Computing},
month = {Nov},
year = {2020},
pages = {41-48},
}

@misc{Siqueira_van_der_Walt_Ushizima_2021,
    author = {de Siqueira, Alexandre Fioravante and van der Walt, Stéfan and Ushizima, Daniela Mayumi},
    title = {Data from: A reusable pipeline for large-scale fiber segmentation on unidirectional fiber beds using fully convolutional neural networks},
    publisher = {Dryad},
    year = {2021},
    month = {jan},
    url = {https://doi.org/10.6078/D1069R},
    doi = {10.6078/D1069R}
}

@article{Blaiszik_2016,
    author = {Blaiszik, B. and Chard, K. and Pruyne, J. and Ananthakrishnan, R. and Tuecke, S. and Foster, I.},
    title = {The Materials Data Facility: Data services to advance materials science research},
    journal = {JOM},
    volume = {68},
    number = {8},
    pages = {2045--2052},
    year = {2016},
    publisher = {Springer US},
    doi = {10.1007/s11837-016-2001-3},
    url = {https://doi.org/10.1007/s11837-016-2001-3}
}

@article{
doi:10.1073/pnas.1812064115,
author = {Kanupriya Pande  and Jeffrey J. Donatelli  and Erik Malmerberg  and Lutz Foucar  and Christoph Bostedt  and Ilme Schlichting  and Petrus H. Zwart },
title = {Ab initio structure determination from experimental fluctuation X-ray scattering data},
journal = {Proceedings of the National Academy of Sciences},
volume = {115},
number = {46},
pages = {11772-11777},
year = {2018},
doi = {10.1073/pnas.1812064115},
URL = {https://www.pnas.org/doi/abs/10.1073/pnas.1812064115},
eprint = {https://www.pnas.org/doi/pdf/10.1073/pnas.1812064115}}

@article{Ushizima2025AscribeND,
  title={Ascribe New Dimensions to Scientific Data Visualization with VR},
  author={Daniela Ushizima and Guilherme Melo dos Santos and Zineb Sordo and Ronald Pandolfi and Jeffrey Donatelli},
  journal={arXiv preprint arXiv:2504.13448},
  year={2025},
  url={https://doi.org/10.48550/arXiv.2504.13448},
  eprint={2504.13448},
  archiveprefix={arXiv},
  primaryclass={cs.GR}
}

@Article{Quenum:2023,
AUTHOR = {Quenum, Jerome and Zenyuk, Iryna V. and Ushizima, Daniela},
TITLE = {Lithium Metal Battery Quality Control via Transformer–CNN Segmentation},
JOURNAL = {Journal of Imaging},
VOLUME = {9},
YEAR = {2023},
NUMBER = {6},
ARTICLE-NUMBER = {111},
URL = {https://www.mdpi.com/2313-433X/9/6/111},
PubMedID = {37367459},
ISSN = {2313-433X},
DOI = {10.3390/jimaging9060111}
}

\end{document}